\documentclass[a4paper]{jpconf}

\usepackage[normalem]{ulem}

\usepackage{braket,bm,subfig, cite}

\usepackage{tikz}
\newcommand{\boxalign}[2][0.97\textwidth]{
 \par\noindent\tikzstyle{mybox} = [draw=blue,inner sep=5pt]
 \begin{center}\begin{tikzpicture}
  \node [mybox] (box){%
   \begin{minipage}{#1}{\vspace{-5mm}#2}\end{minipage}
  };
 \end{tikzpicture}\end{center}
}

\usepackage[pass]{geometry} 
\usepackage{color}
\usepackage{mathrsfs,mathtools, wasysym,amsmath, amsfonts,amssymb,graphicx}
  \DeclareMathAlphabet{\mathpzc}{OT1}{pzc}{m}{it}

\renewcommand{\Re}{\mathrm{Re}}

\newcommand{\gC}{\widetilde{C}}

\renewcommand{\lsim}{\lesssim}

\newcommand{\CP}{C\!P}
\newcommand{\gCP}{\widetilde{C}\!P}

\newcommand{\ve}[1]{\mathbf{#1}}

\newcommand{\DiT}{(2 \pi)^3 \delta^{(3)}}

\newcommand{\Esdu}[4]
{\left[(2 E_{#1}(\ve #2))^{-{1}/{2}}\right]_{#3}^{\phantom{#3}#4}}

\newcommand{\su}[4]{[u(\ve #2, #1)]_{#3}^{\phantom #3 #4}}

\newcommand{\sv}[4]{[v(\ve #2,#1)]_{#3}^{\phantom #3 #4}}

\newcommand{\n}[6]{[n_{#2}^{#1}(\ve #3,#6)]_{#4}^{\phantom{#4}#5}}
\newcommand{\nb}[6]{[\bar{n}_{#2}^{#1}(\ve #3,#6)]_{#4}^{\phantom{#4}#5}}

\renewcommand{\mat}[1]{\bm{#1}}

\newcommand{\edu}[4]{\big[e^{#1 i #2 \cdot x}\big]_{#3}^{\phantom{#3} #4}}

\newcommand{\h}[2]{h_{#1}^{\phantom{#1}#2}}
\newcommand{\hs}[2]{h_{\phantom{#1}#2}^{#1}}
\newcommand{\hr}[2]{\mathbf{h}_{#1}^{\phantom{#1}#2}}
\newcommand{\hrs}[2]{\mathbf{h}_{\phantom{#1}#2}^{#1}}
\newcommand{\hrc}[2]{[\mathbf{h}^{\tilde{c}}]^{#1}_{\phantom{#1}#2}}
\newcommand{\hrcs}[2]{[\mathbf{h}^{\tilde{c}}]^{\phantom{#1}#2}_{#1}}

\newcommand{\Tdu}[5]
{{#1}_{#2 \phantom{#3} #4 \phantom{#5}}^{\phantom{#2} #3 \phantom{#4}  #5}}

\DeclareMathAlphabet{\mathpzc}{OT1}{pzc}{m}{it} 

\newcommand{\D}[2]{\mathrm{d}^{#1}{#2}}

\begin{document}

\title{Flavour effects in Resonant Leptogenesis from semi-classical and Kadanoff-Baym approaches}

\author[a]{P.~S.~Bhupal Dev$^1$, Peter Millington$^2$, Apostolos Pilaftsis$^1$, \underline{Daniele Teresi}$^1$}

\address{$^1$ Consortium for Fundamental Physics,
  School of Physics and Astronomy, \\ 
  University of Manchester, Manchester M13 9PL, United Kingdom}

\address{$^2$ Physik Department T70, James-Franck-Stra\ss e,
Technische Universit\"{a}t M\"{u}nchen, \\ 85748 Garching, Germany}

\ead{Bhupal.Dev@manchester.ac.uk, P.W.Millington@tum.de, Apostolos.Pilaftsis@manchester.ac.uk, Daniele.Teresi@manchester.ac.uk}

\begin{abstract}
Flavour effects play an important role in the statistical evolution of particle number densities in several particle physics phenomena. We present a fully flavour-covariant formalism for transport phenomena, in order to consistently capture all flavour effects in the system. We explicitly study the scenario of Resonant Leptogenesis (RL), and show that flavour covariance requires one to consider generically off-diagonal number densities, rank-4 rate tensors in flavour space, and non-trivial generalization of the discrete symmetries $C$, $P$ and $T$.
The flavour-covariant transport equations, obtained in our semi-classical framework, describe the effects of {\em three} relevant physical phenomena: coherent heavy-neutrino oscillations, quantum decoherence in the charged-lepton sector, and resonant $\CP$ violation due to heavy-neutrino mixing. We show quantitatively that the final asymmetry predicted in RL models may vary by as much as an order of magnitude between partially flavour off-diagonal treatments.
A full field-theoretic treatment in the weakly-resonant regime, based on the Kadanoff-Baym (KB) equations, confirms that heavy-neutrino oscillations and mixing are two {\em distinct} phenomena, and reproduces the results obtained in our semi-classical framework. Finally, we show that the quasi-particle ansaetze, often employed in KB approaches to RL, discard the phenomenon of mixing, capturing only oscillations and leading to an underestimate of the final asymmetry by a factor of order 2.
\end{abstract}

\section{Introduction}\label{sec:intro}
Leptogenesis~\cite{Fukugita:1986hr} is an elegant unifying framework for dynamically generating both the measured matter-antimatter asymmetry in our Universe and the observed smallness of the light neutrino masses~\cite{pdg}. In scenarios of Resonant Leptogenesis (RL)~\cite{Pilaftsis:1997dr, Pilaftsis:2003gt}, this mechanism may be testable in foreseeable laboratory experiments. RL relies on the fact that the $\varepsilon$-type $\CP$-asymmetry becomes dominant~\cite{Flanz:1994yx} and gets resonantly enhanced, when at least two of the heavy  neutrinos have a small mass difference comparable to their  decay  widths~\cite{Pilaftsis:1997dr}. This resonant enhancement allows a successful low-scale leptogenesis~\cite{Pilaftsis:2003gt, Pilaftsis:2005rv},  whilst retaining perfect agreement with the light-neutrino oscillation data. The level of testability is further extended in the scenario of Resonant $\ell$-Genesis (RL$_{\ell}$), where the final lepton asymmetry is dominantly generated and stored in a {\it single} lepton flavour $\ell$~\cite{Pilaftsis:2004xx, Deppisch:2010fr}. In such models, the heavy neutrinos could be as light as the electroweak scale~\cite{Pilaftsis:2005rv}, whilst still having sizable couplings to other charged-lepton flavours $\ell'\neq \ell$. Thus, RL$_{\ell}$ scenarios may be directly testable at the energy frontier in the run-II phase of the LHC~\cite{Deppisch:2015}, as well as in various low-energy experiments searching for lepton flavour/number violation~\cite{Raidal:2008jk} at the intensity frontier.

Flavour effects in both heavy-neutrino and charged-lepton sectors, as well as the interplay between them, can play an important role in determining the final lepton asymmetry in low-scale leptogenesis models (for a review, see e.g.~\cite{Blanchet:2012bk}). These intrinsically-quantum effects  can, in principle, be accounted for by extending the classical  flavour-diagonal Boltzmann equations for the number  densities of individual flavour species to a  semi-classical evolution equation for a
{\it matrix  of  number densities}, analogous to the formalism presented in~\cite{Sigl:1993} for light neutrinos. This so-called `density matrix' formalism has been adopted to describe flavour effects in various leptogenesis scenarios~\cite{Abada:2006fw, Nardi:2006fx, Akhmedov:1998qx}. It was recently shown~\cite{Dev:2014laa}, in a semi-classical approach, that a consistent treatment of {\em all} pertinent flavour effects, including flavour mixing, oscillations and off-diagonal (de)coherences, necessitates a {\em fully} flavour-covariant formalism, in order to provide a complete and unified description of RL; for a summary, see~\cite{Dev:2014tpa}. In this flavour-covariant formalism, the resonant mixing of different heavy-neutrino flavours and coherent oscillations between them are found to be two {\em distinct} physical phenomena, in analogy with the experimentally-distinguishable phenomena of mixing and oscillations in the neutral $K$-, $D$-, $B$- and $B_s$-meson systems~\cite{pdg}.

One can go beyond the semi-classical `density-matrix' approach to leptogenesis by means of a quantum field-theoretic analogue of the Boltzmann equations, known as the Kadanoff-Baym (KB) equations~\cite{Baym:1961zz} (for a review, see e.g.~\cite{Berges:2004yj}). Such `first-principles' approaches to leptogenesis~\cite{Buchmuller:2000nd} are, in principle, capable of accounting consistently for all flavour effects, in addition to off-shell and finite-width effects, including thermal corrections. However, it is often necessary to use truncated gradient expansions and quasi-particle ansaetze to relate the propagators appearing in the KB equations to particle number densities. Recently, using the novel perturbative formulation of thermal field theory developed in~\cite{Millington:2012pf}, it was shown~\cite{Dev:2014wsa} that quantum transport equations for leptogenesis can be obtained from the KB formalism without the need for gradient expansion or quasi-particle ansaetze, thereby capturing fully the pertinent flavour effects. Specifically, the source term for the lepton asymmetry obtained, at leading order, in this KB approach~\cite{Dev:2014wsa} was found to be exactly the same as that obtained in the semi-classical flavour-covariant approach of~\cite{Dev:2014laa}, confirming that flavour mixing and oscillations are indeed two {\em physically-distinct} phenomena. The proper treatment of these flavour effects may have a significant effect upon the final lepton asymmetry, as compared to partially flavour-dependent or flavour-diagonal limits, thereby altering the viable parameter space for models of RL and impacting upon the prospects of testing the leptogenesis mechanism. 

The plan of these proceedings is as follows. In Section~\ref{sec:covariant}, we review the main features of our {\it fully}  flavour-covariant formalism in the context of leptogenesis. 
In Section~\ref{sec:semiclassical}, we present the Markovian flavour-covariant transport
equations for lepton and heavy-neutrino  number densities with arbitrary flavour content. We also discuss a numerical example to illustrate the full impact of the flavour off-diagonal effects within the context of an RL$_\tau$ model. In Section~\ref{sec:quantum}, we derive the quantum transport equations relevant to the source term for the lepton asymmetry, following a well-defined perturbative loopwise truncation scheme, making comparison with the semi-classical approach discussed in Section~\ref{sec:semiclassical}. Our conclusions are given in Section~\ref{sec:conclusion}.   
 
\section{Flavour-covariant formalism}\label{sec:covariant}

We consider the lepton-doublet field operators $L_l$ (with $l=1, 2,  \dots,  \mathcal{N}_{L}$) and right-handed Majorana neutrino field operators $N_{\rm R, \alpha}$ (with $\alpha=1, 2, \dots, \mathcal{N}_N$), with arbitrary flavour content, transforming as follows in the fundamental representation of $U(\mathcal{N}_{L})\otimes U(\mathcal{N}_{N})$: 
\begin{subequations}
\begin{gather}
\hspace{-1.0em}L_l \ \to \ L'_l \ = \ V_l^{\phantom{l}m}L_m\;,\qquad 
L^l \ \equiv \ (L_l)^{\dag} \ \to \ L'^l \ = \ V^l_{\phantom{l}m}L^m\;,
\\
\hspace{-1.0em}
N_{\mathrm{R},\, \alpha} \ \to \ N'_{\rm R, \alpha} \ = \ U_{\alpha}^{\phantom{\alpha}\beta}N_{\mathrm{R},\,\beta},\qquad 
N_{\mathrm{R}}^{\alpha} \ \equiv \ (N_{\mathrm{R},\, \alpha})^\dag \ \to \ 
N_{\rm R}'^{ \alpha} \ =  \ U^{\alpha}_{\phantom{\alpha}\beta}N_{\mathrm{R}}^{\beta}\; ,
\end{gather}
\end{subequations}
where $V_l^{\phantom{l}m} \in U(\mathcal{N}_{L})$ and $U_{\alpha}^{\phantom{\alpha}\beta} \in U(\mathcal{N}_{N})$. The relevant neutrino Lagrangian is given by 
\begin{equation}
  -\mathcal{L}_N  \  = \  \h{l}{\alpha}  \overline L^{l} 
  \widetilde{\Phi}  N_{\rm R, \alpha} 
  + \frac{1}{2} \overline{N}_{\rm R, \alpha}^C  [M_N]^{\alpha \beta} 
  N_{\rm R, \beta} + {\rm H.c.}\;,
  \label{eq:L}
\end{equation}
where $\widetilde{\Phi}=i\sigma_2\Phi^*$ is the isospin conjugate of the Higgs doublet $\Phi$ and the superscript $C$ denotes charge conjugation. The Lagrangian~\eqref{eq:L} transforms covariantly under $U(\mathcal{N}_{L})\otimes U(\mathcal{N}_{N})$, provided the Yukawa couplings and Majorana mass matrix transform as
\begin{align}
  \h{l}{\alpha} \ \rightarrow \ h_{l}'^{\ \alpha} \ = \ V_l^{\phantom l m}
  \;
  U^\alpha_{\phantom{\alpha} \beta} \; \h{m}{\beta} \; ,
\qquad   [M_N]^{\alpha \beta} \ \rightarrow \ 
[M'_N]^{\alpha \beta} \ = \ U^\alpha_{\phantom{\alpha} \gamma} \;
  U^\beta_{\phantom{\beta} \delta} \; [M_N]^{\gamma \delta} \;.
\end{align}
In this flavour-covariant formalism, the plane-wave decompositions of the field operators are written in a manifestly flavour-covariant way, e.g.
\begin{align}
L_l(x) \ & = \  \sum_{s=+,-} \int_{\ve p} \Esdu{L}{p}{l}{i}
\notag \\ &\qquad \times \Big( \edu{-}{p}{i}{j} \su{s}{p}{j}{k} \,
    b_k(\ve p,s,0)  \; +\: \edu{}{p}{i}{j} \, \sv{s}{p}{j}{k} \,
    d_k^{\dagger}(\ve p,s,0)
  \Big)\;,
\label{eq:Lfield}
\end{align}
where $\int_{\mathbf{p}}\equiv\int\!\frac{\D{3}{\mathbf{p}}}{(2\pi)^3}$, $s$ is the helicity index and $[E_L^2(\mathbf{p})]_{l}^{\phantom{l}m}=\mathbf{p}^2\delta_{l}^{\phantom{l}m}+
[M_L^{\dag}M_L]_{l}^{\phantom{l}m}$, with $M_L$ being the charged-lepton mass matrix. 
Thus, flavour covariance requires the Dirac four-spinors $\su{s}{p}{j}{k}$ and $\sv{s}{p}{j}{k}$ to transform as rank-$2$ tensors in flavour space. The creation and annihilation operators $b^k\equiv b_k^{\dag}$, $b_k$, $d^{\dagger,\, k}\equiv d_k$ and $d_k^{\dagger}$ satisfy the equal-time anti-commutation relations
\begin{align}
  \label{eq:b_d_anticomm}
 \big\{ b_l(\ve p,s,\tilde{t}), \,
  b^{m}(\ve p',s',\tilde{t}) \big\}  \  &= \ \big\{d^{\dagger, m}(\ve p,s,\tilde{t}) , \,
  d_l^{\dagger}(\ve p',s',\tilde{t}) \big\} 
 \ = \ \DiT{(\ve p - \ve p')} \, \delta_{s s'}\,
  \delta_l^{\phantom l m}\;.
\end{align}

For the heavy {\em Majorana} neutrino creation and annihilation operators $a^{\alpha}(\ve k,r,\tilde{t})$ and $a_{\alpha}(\ve k,r,\tilde{t})$, it is necessary to introduce the flavour-covariant Majorana constraint
\begin{equation}
  d^{\dagger , \alpha}(\ve k,-\,r,\tilde{t})\ 
  = \ G^{\alpha \beta} \, b_\beta(\ve k,r,\tilde{t}) \ \equiv\ G^{\alpha\beta}a_{\beta}(\ve k,r,\tilde{t}) \;,
  \label{eq:def_G}
\end{equation}
where $G^{\alpha \beta}\equiv[ U^* U^\dagger ]^{\alpha \beta}$ are the elements of a unitary matrix $\bm{G}$, which transforms as a contravariant rank-$2$ tensor under $U(\mathcal{N}_N)$. Notice that $\bm{G}=\bm{1}$ in the mass eigenbasis.

Similar flavour rotations are {\it forced} by flavour-covariance under the discrete symmetry transformations $C,~P$ and $T$. This necessarily leads to {\it generalized} $C$ and $T$ transformations:
\begin{subequations} 
\begin{align}
 b_l(\ve p,s,\tilde{t})^{\widetilde{C}} \ & \equiv \ \mathcal{G}^{lm} \,
  b_m(\ve p,s,\tilde{t})^{C} \
  = \ -i \, d^{\dagger,l}(\ve p,s,\tilde{t}) \; , \label{gC}\\
b_l(\ve p,s,\tilde{t})^P \ & 
  = \ - s \, b_l(-\ve p,-s,\tilde{t})\;, \label{gP} \\
 b_l(\ve p,s,\tilde{t})^{\widetilde{T}} \ &\equiv \
  \mathcal{G}_{lm} \, b_m(\ve p,s,\tilde{t})^{T} \
  = \   b_l(-\ve p,s,-\tilde{t}) \;, \label{gT}
\end{align}
\end{subequations}
where $\mathcal{G}^{lm}\equiv [V^*V^\dag]^{lm}$ is the lepton analogue of the heavy-neutrino tensor $\bm{G}$.

We may now define the matrix number densities of the leptons and heavy neutrinos, which describe completely the flavour content of the system: 
\begin{subequations}
\begin{align}
  \hspace{-0.85em}\n{L}{s_1 s_2}{p}{l}{m}{t} \ & \equiv \ \mathcal V_3^{-1}
  \langle b^m(\ve p, s_2,\tilde{t})
  b_l(\ve p, s_1,\tilde{t})\rangle_t \;,
  \label{eq:def_n_1}\\
  \hspace{-0.85em}\nb{L}{s_1 s_2}{p}{l}{m}{t} \ & \equiv \ \mathcal V_3^{-1}
  \langle d_l^{\dagger}(\ve p,s_1,\tilde{t})
  d^{\dagger,m}(\ve p,s_2,\tilde{t})\rangle_t \;,
  \label{eq:def_n_2}\\
  \hspace{-0.85em}\n{N}{r_1 r_2}{k}{\alpha}{\beta}{t} \ & \equiv \ 
  \mathcal V_3^{-1}
  \langle a^{\beta}(\ve k,r_2,\tilde{t})
  a_\alpha(\ve k,r_1,\tilde{t})\rangle_t \;.
  \label{eq:def_n_3} 
\end{align}
\end{subequations}
Here,  $\mathcal V_3$ is the infinite coordinate three-volume and $t=\tilde{t}-\tilde{t}_i$ is the macroscopic  time, equal to the time  interval between specification of initial  conditions ($\tilde{t}_i$)
and subsequent observation  of  the  system ($\tilde{t}$).   
The total number densities $\bm{n}^X(t)$ are obtained by tracing over helicity and isospin indices and integrating over the three-momenta. For the Majorana neutrinos, $\bm{n}^N$ and $\overline{\bm{n}}^N$ are not independent quantities and are related by the generalized Majorana constraint~\eqref{eq:def_G}. 

Using the generalized discrete transformations~\eqref{gC}--\eqref{gT}, we can define the flavour-covariant $\widetilde{C}P$-``even'' and -``odd'' quantities 
\begin{equation}
\qquad \underline{\bm{n}}^N\ = \ \frac{1}{2}\Big(\bm{n}^N\:+\:\overline{\bm{n}}^N\Big)\;,\qquad \bm{\delta n}^N\ =\ \bm{n}^N\:-\:\overline{\bm{n}}^N\;, \qquad \bm{\delta n}^L\ = \ \bm{n}^L\:-\:\overline{\bm{n}}^L\;,  
\end{equation}
which will be used in the following section to write down the flavour-covariant rate equations. 

\section{Flavour-covariant semi-classical rate equations}
\label{sec:semiclassical}

We first derive a master equation governing the time evolution of the matrix number densities $\mat{n}^X(\ve p,t)$, as given in~\eqref{eq:def_n_1}--\eqref{eq:def_n_3}. By using the Liouville-von Neumann and Heisenberg equations of motion and subsequently performing a Wigner-Weisskopf approximation in the Markovian limit~\cite{Dev:2014laa}, we find
\begin{equation}
  \label{eq:master}
  \frac{\D{}{}}{\D{}{t}} \mat{n}^X(\ve k, t) \
  \simeq \ i  \langle \, [H_0^X,\  \mat{\check{n}}^{X}(\ve k, t) ] \,
  \rangle_t  \ -\:\frac{1}{2} \int_{-\infty}^{+\infty} \D{}{t'} \;
  \langle \, [H_{\rm int}(t'),\
  [H_{\rm int}(t),\ \mat{\check{n}}^{X}(\ve k, t)]] \, \rangle_{t} \; ,
\end{equation}
where $H^X_0$ and $H_{\rm int}$ are respectively the free and interaction parts of the Hamiltonian and 
\begin{equation}
  \label{eq:def_n_rho}
  \mat{n}^{X}(\ve k, t) \ \equiv \ \langle
  \mat{\check{n}}^{X}(\ve k,\tilde{t};\tilde{t}_i) \rangle_t 
  \ = \ \Tr\left\{\rho(\tilde{t};\tilde{t}_i) \,
    \mat{\check{n}}^{X}(\ve k, \tilde{t};\tilde{t}_i) \right\}
\end{equation}
is the ensemble expectation value of the quantum-mechanical number-density operator $\mat{\check{n}}^X(\ve k,\tilde{t};\tilde{t}_i)$, 
in which $\rho(\tilde{t};\tilde{t}_i)$ is the interaction-picture density operator. The first term on the RHS of~\eqref{eq:master}, involving the free Hamiltonian, is responsible for flavour oscillations, whereas the second term contains the collision terms of the generalized Boltzmann equations. Explicitly, for the system of charged leptons and heavy neutrinos, we find~\cite{Dev:2014laa}
\begin{subequations}
\begin{align}
  &\frac{\D{}{} }{\D{}{t}} \, \n{L}{s_1 s_2}{p}{l}{m}{t} 
  =  -  i 
  \Big[{E}_L(\ve p), \,{n}^{L}_{s_1 s_2}(\ve p,t)
  \Big]_{l}^{\phantom{l}m} +  [{C}^L_{s_1 s_2}(\ve p,t)]_l^{\phantom l m}\;,
  \label{eq:evol_lept}\\
  &\frac{\D{}{} }{\D{}{t}} \,\n{N}{r_1 r_2}{k}{\alpha}{\beta}{t}  
   =  -  i \Big[{E}_N(\ve k), \,
  {n}^{N}_{r_1 r_2}(\ve k,t)\Big]_{\alpha}^{\phantom{\alpha}\beta}\!
  +  [{C}^{N}_{r_1 r_2}(\ve k,t)]_\alpha^{\phantom \alpha \beta} 
+  G_{\alpha \lambda} \,
  [\overline{{C}}^N_{r_2 r_1}(\ve k,t)]_{\mu}^{\phantom{\mu} \lambda} \,
  G^{\mu \beta}\;.  
  \label{eq:evol_neu}
\end{align}
\end{subequations}
The collision terms $[{C}^L_{s_1 s_2}(\ve p,t)]_l^{\phantom l m}$ and $[{C}^{N}_{r_1 r_2}(\ve k,t)]_\alpha^{\phantom \alpha \beta}$ involve the product of two new {\em rank-4} tensors in flavour space, namely, the statistical number density tensor and the absorptive rate tensor, whose appearance is necessary for the flavour covariance of the formalism~\cite{Dev:2014laa}. The emergence of these rank-4 tensors may be understood in terms of the unitarity cuts of the partial self-energies, as was shown by an explicit calculation of the relevant transition amplitudes using a generalized optical theorem in~\cite{Dev:2014laa}. The off-diagonal components of the rate tensor are responsible for the evolution of flavour-coherences in the system. 

In  the limit  when two (or more) of the heavy Majorana neutrinos become  degenerate, the $\varepsilon$-type  $ \CP$-violation can be resonantly enhanced, even up to order one~\cite{Pilaftsis:1997dr}, due to the interference between the tree-level and self-energy-corrected decays. 
In this regime, finite-order perturbation  theory breaks  down and  one must resum the  self-energy corrections in order to account for the heavy-neutrino mixing effects. In the semi-classical approach,  we perform such resummation in an effective way by replacing the  tree-level  neutrino  Yukawa couplings $\hs{l}{\alpha}$ by  their resummed  counterparts $\hr{l}{\alpha}$ in the  transport equations.  In the next section, this approach will be justified for the weakly-resonant regime of RL, where $\Gamma_{N_{\alpha,\beta}} < |m_{N_\alpha} - m_{N_\beta}| \ll m_{N_{\alpha,\beta}}$, by using a `first-principles' field-theoretic approach. The explicit algebraic form of the resummed  neutrino Yukawa  couplings in  the heavy-neutrino mass eigenbasis can be found in~\cite{Pilaftsis:2003gt}; the corresponding form in a general flavour basis may be obtained by
the   appropriate  flavour  transformation,   i.e.~$\hr{l}{\alpha}  =
V_{l}^{\ m}U^{\alpha}_{\  \beta}\widehat{\mathbf{h}}_{m}^{\ \ \beta}$,
where                  $\widehat{\mathbf{h}}_{m}^{\                  \
\beta}\equiv\widehat{\mathbf{h}}_{m\beta}$ in the mass eigenbasis~\cite{Dev:2014laa}. 

In addition, we make the following reasonable approximations to simplify the flavour-covariant rate equations for RL: we (i) assume kinetic equilibrium, which is ensured by the presence of fast elastic-scattering processes; (ii) work in the classical-statistical regime; (iii) neglect thermal and chemical-potential effects~\cite{Pilaftsis:2005rv}; and (iv) neglect the  mass  splitting  between different heavy-neutrino flavours inside thermal integrals, using an average  mass $m_N$ and energy $E_N(\ve k) = (|\ve  k|^2 + m_N^2)^{1/2}$, as is appropriate since the average momentum scale $|\ve k|\sim T \gg |m_{N_\alpha}-m_{N_\beta}|$. In order to guarantee the correct equilibrium behaviour, we must also include the effect of the  $\Delta L=0$ and $\Delta L=2$ scattering processes, with proper real intermediate state (RIS) subtraction~\cite{Kolb:1980, Pilaftsis:2003gt, Dev:2014laa}.  As detailed in~\cite{Dev:2014laa}, it is necessary to account for thermal corrections in the RIS contributions, when considering off-diagonal heavy-neutrino flavour correlations. Finally, it is important to include the effect of the charged-lepton Yukawa couplings, which are responsible for the decoherence of the charged leptons towards their would-be mass eigenbasis. 

Taking into account the expansion of the Universe, we then obtain the following  {\it manifestly} flavour-covariant rate equations for the normalized $\gCP$-``even" number 
density matrix $\mat{\underline{\eta}}^{N}$ and $\gCP$-``odd" number density 
matrices $\mat{\delta \eta}^N$ and $\mat{\delta \eta}^L$ (where $\eta^X=n^X/n^{\gamma}$, with $n^\gamma$ being the photon number density)~\cite{Dev:2014laa}:
\vspace{0.5cm}
\begin{subequations}
\boxalign[0.97\textwidth]{
\begin{align}
 \frac{H_{N} \, n^\gamma}{z}\,
   \frac{\D{}{[\underline{\eta}^{N}]_{\alpha}^{\phantom{\alpha}\beta}}}{\D{}{z}} \   &= \ - \, i \, \frac{n^\gamma}{2} \,
  \Big[\mathcal{E}_N,\, \delta \eta^{N}\Big]_\alpha^{\phantom \alpha \beta}
+ \, \Tdu{\big[\widetilde{\rm Re}
    (\gamma^{N}_{L \Phi})\big]}{}{}{\alpha}{\beta} \,
  - \, \frac{1}{2 \, \eta^N_{\rm eq}} \,
  \Big\{\underline{\eta}^N, \, \widetilde{\rm Re}(\gamma^{N}_{L \Phi})
  \Big\}_{\alpha}^{\phantom{\alpha}\beta} \;,
  \label{eq:evofinal2}\\[6pt]
  \frac{H_{N} \, n^\gamma}{z}\,
  \frac{\D{}{[\delta \eta^N]_\alpha^{\phantom \alpha \beta}}}{\D{}{z}} \ 
  &= \ - \, 2 \, i \, n^\gamma \,
  \Big[\mathcal{E}_N,\, \underline{\eta}^{N}\Big]_\alpha^{\phantom \alpha \beta} \, + \, 2\, i\,  \Tdu{\big[\widetilde{\rm Im}
    (\delta \gamma^{N}_{L \Phi})\big]}{}{}{\alpha}{\beta} \, - \, 
  \frac{i}{\eta^N_{\rm eq}} \, \Big\{\underline{\eta}^N, \,
  \widetilde{\rm Im}
  (\delta\gamma^{N}_{L \Phi}) \Big\}_{\alpha}^{\phantom{\alpha}\beta} \notag\\
  &\quad\;\; - \, \frac{1}{2 \, \eta^N_{\rm eq}}  \,
  \Big\{\delta \eta^N, \, \widetilde{\rm Re}(\gamma^{N}_{L \Phi})
  \Big\}_{\alpha}^{\phantom{\alpha}\beta}
  \label{eq:evofinal3}\;, \\[6pt]
 \frac{H_{N} \, n^\gamma}{z}\, 
\frac{\D{}{[\delta \eta^L]_l^{\phantom l m}}}
  {\D{}{z}} \ 
  &= \ - \, \Tdu{[\delta \gamma^{N}_{L \Phi}]}{l}{m}{}{} \,
  +\, \frac{[\underline{\eta}^{N}]_{\beta}^{\phantom{\beta}\alpha}}
  {\eta^N_{\rm eq}} \,
  \Tdu{[\delta \gamma^{N}_{L \Phi}]}{l}{m}{\alpha}{\beta} 
   + \, \frac{[\delta \eta^N]_{\beta}^{\phantom\beta \alpha}}{2\,\eta^N_{\rm eq}}  \,
  \Tdu{[\gamma^{N}_{L \Phi}]}{l}{m}{\alpha}{\beta} \notag\\ 
  &\quad\;\; - \frac{1}{3} \,
  \Big\{ \delta {\eta}^{L} , \,
  {\gamma}^{L\Phi}_{L^{\tilde{c}} \Phi^{\tilde{c}}} 
  + {\gamma}^{L\Phi}_{L \Phi}\Big\}_{l}^{\phantom l m}  
  \, - \, \frac{2}{3} \, \Tdu{[\delta {\eta}^L]}{k}{n}{}{} \,
  \Tdu{[{\gamma}^{L\Phi}_{L^{\tilde{c}} \Phi^{\tilde{c}}} - {\gamma}^{L\Phi}_{L \Phi}]}{n}{k}{l}{m} 
 \notag\\[3pt]
  & \quad\;\; - \frac{2}{3} \, 
  \Big\{\delta \eta^L, \, 
  \gamma_{\rm dec } \Big\}_{l}^{\phantom l m} \,
  +\, [\delta \gamma_{\rm dec}^{\rm back}]_{l}^{\phantom l m}\;.
  \label{eq:evofinal1}
\end{align}}
\end{subequations}
\vspace{0.5cm}
Here, $H_N$ is the Hubble parameter at $z\equiv m_N/T=1$ and we have defined the thermally-averaged heavy-neutrino energy matrix
\begin{equation}
\label{thermen}
\mat{\mathcal{E}}_{\!N} \ = \ \frac{g_N}{n^N_{\rm eq}} \, \int_{\ve k} \, \mat{E}_N(\ve k) \, e^{-E_N(\ve k)/T} \;,
\end{equation}
where $g_N=2$ counts the helicity degrees of freedom. In addition, $\mat\gamma^N_{L\Phi}$ and $\mat{\delta\gamma}^N_{L\Phi}$ are respectively the $\gCP$-``even" and -``odd" thermally-averaged rate tensors, describing heavy-neutrino decays and inverse decays and written in terms of the resummed Yukawa couplings as
\begin{equation}
[\gamma^{N}_{L \Phi}]_{\alpha}^{\phantom{\alpha}\beta} \ \equiv \ \int_{N L \Phi}  \big( \hrs{}{\alpha} \hr{}{\beta} \, +\, \hrc{}{\alpha} \hrcs{}{\beta}\big) \;,\qquad 
[\delta \gamma^{N}_{L \Phi}]_{\alpha}^{\phantom{\alpha}\beta} \ \equiv \ \int_{N L \Phi}  \big( \hrs{}{\alpha} \hr{}{\beta} \, -\, \hrc{}{\alpha} \hrcs{}{\beta}\big) \;,
\end{equation}
where $\tilde{c}$ denotes the generalized $\CP$-conjugate and we have used the shorthand notation
\begin{equation}
\int_{NL\Phi} \ \equiv \ \int \D{}{\Pi_N}  \int \D{}{\Pi_L} \int \D{}{\Pi_\Phi} \, (2 \pi)^4 \, \delta^{(4)}(p_N-p_L-p_\Phi) \, e^{-p_N^0/T} \;,
\end{equation}
with the phase-space measure for the species $X$ given by
\begin{equation}
\D{}{\Pi}_X \ \equiv \ \frac{\D{4}{p_X}}{(2 \pi)^4} \, 2\pi \delta(p_X^2 - M_X^2) \, \theta(p_X^0) \;.
\end{equation} 
In~\eqref{eq:evofinal1}, $\mat\gamma^{L\Phi}_{L \Phi}$ and $\mat\gamma^{L\Phi}_{L^{\tilde{c}} \Phi^{\tilde{c}}}$ respectively describe the washout due to $\Delta L = 0$ and $\Delta L = 2$ resonant scattering, and $\mat\gamma_{\rm dec}$ and $\mat{\delta \gamma}_{\rm dec}^{\rm back}$ govern the charged-lepton decoherence~\cite{Dev:2014laa}. Lastly, for a Hermitian matrix  $\mat{A}$, we have defined the flavour-covariant generalized real  and
imaginary parts
\begin{align}
  \big[\widetilde{\rm Re}(A)\big]_{\alpha}^{\phantom{\alpha}\beta} \ 
  \equiv \ \frac{1}{2} \, \Big( \Tdu{A}{\alpha}{\beta}{}{}  
  +  G_{\alpha \lambda} \,\Tdu{A}{\mu}{\lambda}{}{}\,
  G^{\mu \beta}\Big)\;, \qquad
  \big[\widetilde{\rm Im}(A)\big]_{\alpha}^{\phantom{\alpha}\beta} \ 
  \equiv \ \frac{1}{2 i} \, 
  \Big( \Tdu{A}{\alpha}{\beta}{}{}  
  -  G_{\alpha \lambda} \,\Tdu{A}{\mu}{\lambda}{}{}\,
  G^{\mu \beta}\Big)\;, \label{4.27}
\end{align}
which reduce to the usual real and imaginary parts in the heavy-neutrino mass eigenbasis.

The             flavour-covariant            rate            equations~\eqref{eq:evofinal2}--\eqref{eq:evofinal1} provide a complete and unified  description of the generation of the lepton asymmetry in RL,  consistently describing  the following {\it physically-distinct} effects in a  single framework: 
\begin{itemize}
\item [(i)]  Lepton asymmetry due to the $\CP$-violating {\it resonant mixing}  between heavy neutrinos, as described by the $\gCP$-``odd'' rates $\mat{\delta\gamma}^N_{L\Phi}$, appearing in the first two terms on the RHS of~\eqref{eq:evofinal1}. 
This provides a 
flavour-covariant generalization of the mixing effects discussed earlier in~\cite{Pilaftsis:2003gt}. 

\item [(ii)] Lepton asymmetry via coherent heavy-neutrino {\it oscillations}. This is an ${\cal O}(h^4)$ effect on the {\it total} lepton asymmetry~\cite{Dev:2014laa}, thus differing from the ${\cal O}(h^6)$ mechanism studied in~\cite{Akhmedov:1998qx}, which typically takes place at temperatures much higher than the sterile neutrino masses. 

\item [(iii)] {\it Decoherence} effects  due to charged-lepton Yukawa couplings. Our description of these effects generalizes the analysis in~\cite{Abada:2006fw} to an arbitrary flavour basis. 
\end{itemize}

In order to illustrate the importance of the flavour effects captured {\it only} by the flavour-covariant rate equations~\eqref{eq:evofinal2}--\eqref{eq:evofinal1}, we consider an
RL$_\tau$  model, comprising an approximately $SO(3)$-symmetric heavy-neutrino sector at the grand unification scale $\mu_X\sim 2\times  10^{16}$ GeV, with masses~$\bm{M}_N(\mu_X)=m_N\bm{1}_3+\bm{\Delta M}_N$~\cite{Pilaftsis:2005rv}. The soft $SO(3)$-breaking mass term is taken to be of the form $\bm{\Delta M}_N=\mathrm{diag}\,(\Delta M_1,\Delta M_2/2,-\,\Delta M_2/2)$. At the electroweak scale, an additional mass splitting arises from the RG running, such that $\bm{M}_N(m_N)=m_N\bm{1}_3+\bm{\Delta M}_N+\bm{\Delta M}_N^{\mathrm{RG}}$. In order to accommodate the smallness of the light neutrino masses in a technically natural manner, we also require the heavy-neutrino Yukawa sector to possess an approximate leptonic $U(1)_{l}$ symmetry. This results in the following structure for the heavy-neutrino Yukawa couplings:
\begin{eqnarray}
  \mat{h} \ \equiv  \ \mat{h}_0+ \mat{\delta h} \ = \ \left(\begin{array}{ccc}
      0 & ae^{-i\pi/4} & ae^{i\pi/4}\\
      0 & be^{-i\pi/4} & be^{i\pi/4}\\
      0 & ce^{-i\pi/4} & ce^{i\pi/4}
    \end{array}\right) \: + \: \left(\begin{array}{ccc}
      \epsilon_e & 0 & 0\\
      \epsilon_\mu & 0 & 0\\
      \epsilon_\tau & 0 &
      0
    \end{array}\right)  \; ,
  \label{yuk}
\end{eqnarray}
where $a,b,c$ are arbitrary complex parameters. If the theory were to have an exact $U(1)_{l}$ symmetry, i.e.~if the $U(1)_{l}$-breaking parameters $\epsilon_{e,\mu,\tau}$ were set to zero, the light neutrinos would remain massless to all orders in perturbation theory. For electroweak-scale heavy neutrinos, we require $|a|,|b|	\lsim 10^{-2}$, in order to be consistent with current light-neutrino mass bounds, and $|c| \lsim 10^{-5}$ and $|\epsilon_{e,\mu,\tau}|\lsim 10^{-6}$, in order to protect the $\tau$ lepton asymmetry from wash-out effects.

\begin{figure}[t]
\centering
\newsavebox{\tempbox}
\sbox{\tempbox}{\includegraphics[scale=0.5]{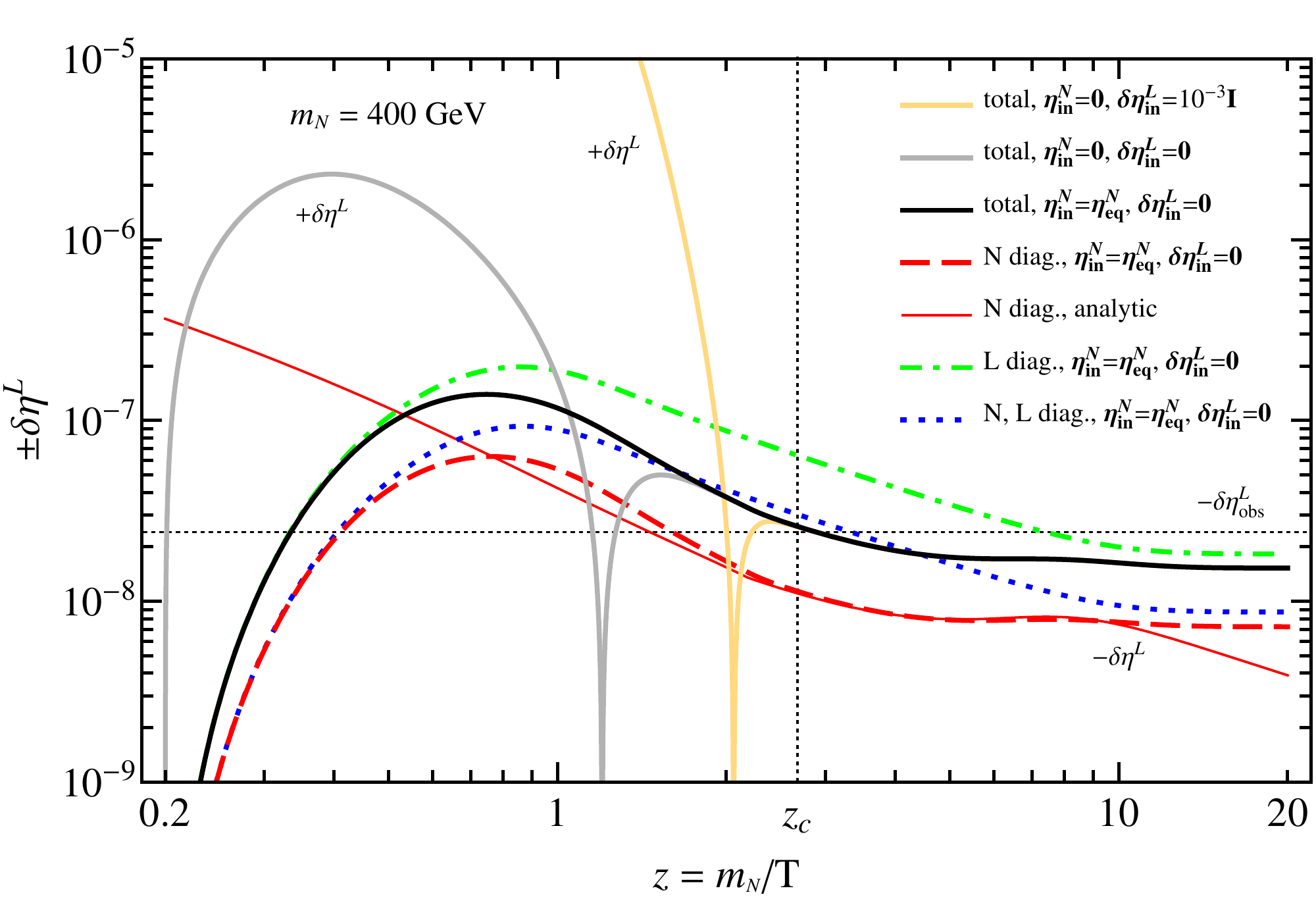}}
\subfloat[ \label{fig1a}]{\includegraphics[scale=0.5]{mN400.pdf}} \hspace{0.2cm}
\subfloat[ \label{fig1b}]{\vbox to \ht\tempbox {\hsize=14em \vfil
\small {\begin{tabular}[b]{l l}\br
Param. & Value \\ \mr
      $m_N$ & 400 GeV \\
      $c$ & $2\times 10^{-7}$\\[2pt]
      $\frac{\Delta M_1}{m_N}$ & $-3\times 10^{-5}$\\[5pt]
      $\frac{\Delta M_2}{m_N}$ & $(-1.21+0.10\,i)\times 10^{-9}$\\
      \mr
      $a$ & 
      $(4.93 - 2.32 \, i) \times 10^{-3}$ \\
      $b$ &
      $(8.04 - 3.79 \, i) \times 10^{-3}$ \\
      $\epsilon_e$ &
      $5.73\,i\times 10^{-8}$ \\
      $\epsilon_\mu$ &
      $4.30\,i\times 10^{-7}$ \\
      $\epsilon_\tau$ &
      $6.39\,i\times 10^{-7}$ \\  
      \br
    \end{tabular}}\vfil}}
\caption{(a) The total lepton asymmetry for a minimal RL$_\tau$  model with the parameters in (b). The thick solid lines show the total asymmetry obtained in our fully flavour-covariant formalism for different intial conditions; the dashed lines show that obtained in various flavour-diagonal limits.}\label{fig1}
\end{figure}

Using the Yukawa coupling given by~\eqref{yuk}, we solve the rate equations~\eqref{eq:evofinal2}--\eqref{eq:evofinal1} numerically to obtain the total lepton asymmetry  $\delta  \eta^L  \equiv  {\rm  Tr}(\mat{\delta  \eta}^L)$ in our flavour-covariant formalism. This is shown in Figure~\ref{fig1a} for a typical set of benchmark values for the Yukawa coupling parameters, as given in Figure~\ref{fig1b}, which is consistent with all current experimental constraints~\cite{Dev:2014laa}. In Figure~\ref{fig1a}, the horizontal dotted line 
shows the value of $\delta\eta^L$ required
to  explain  the  observed  baryon  asymmetry, whereas the vertical line shows the critical temperature $z_c=m_N/T_c$, beyond which the electroweak sphaleron processes become ineffective in converting lepton asymmetry to baryon asymmetry. The thick solid lines show the evolution of $\delta\eta^L$ for three  different initial  conditions, to which the final lepton asymmetry $\delta
\eta^L (z\gg 1)$ is insensitive as a
general consequence of the RL mechanism in the strong-washout regime~\cite{Pilaftsis:2005rv}.  For comparison, Figure~\ref{fig1a} also shows various flavour-diagonal limits, i.e.~when either  the heavy-neutrino (dashed  line)  or the lepton (dash-dotted line) number  density   or both (dotted  line) are diagonal in  flavour space. Also shown (thin solid line) is the approximate analytic solution obtained in~\cite{Dev:2014laa} for 
the case of a diagonal heavy-neutrino number density.  
The enhancement of the lepton asymmetry in the {\it fully} flavour-covariant  formalism (solid line), as compared to assuming a flavour-diagonal heavy-neutrino number density (dashed line), is mainly 
due to coherent  oscillations  between  the heavy-neutrino
flavours,  leading to a factor of  2 increase. Finally, we observe that the predicted lepton asymmetry differs by approximately an order of magnitude between the two partially flavour off-diagonal treatments (dashed and dash-dotted lines). This provides a striking illustration of the importance of capturing all pertinent flavour effects and their interplay by means of a fully flavour-covariant formulation for transport phenomena.

\section{Flavour-covariant quantum transport equations}\label{sec:quantum}

The semi-classical approach detailed in the preceding section has the advantage that it is constructed with physical observables, i.e.~particle number densities, in mind. However, it has the disadvantage that quantum effects, such as finite particle widths, must be incorporated in an effective manner. For instance, in the latter example, we must subtract the RIS contributions from the collision terms~\cite{Kolb:1980}, which would otherwise lead to double counting of decay and inverse-decay processes. It is therefore desirable to seek a more first-principles description of transport phenomena, in which quantum effects are incorporated consistently from the out-set.

Such a description is provided by the Kadanoff-Baym (KB) approach~\cite{Baym:1961zz} (see also~\cite{Prokopec:2003pj, Berges:2004yj}), constructed within the Schwinger-Keldysh closed-time path (CTP) formalism of thermal field theory~\cite{Schwinger:1961}. Therein, one arrives at systems of KB equations by partially inverting the Schwinger-Dyson equation of the 2PI CJT effective action~\cite{Cornwall:1974vz}. Unfortunately, the KB equations describe the spacetime evolution of propagators and, as a result, it is necessary to use approximation schemes in order to obtain the quantum transport equations of particle number densities.

In order to avoid the technical complications of spinor fields, we will consider a scalar model of RL, see~\cite{Dev:2014wsa}, comprising: two real scalar fields $N_{\alpha}$ ($\alpha=1,2$), modelling heavy-neutrinos of two flavours; one complex scalar field $L$, modelling charged-leptons of a single flavour; and a real scalar field $\Phi$, modelling the Standard Model Higgs. The lepton number can be associated with the global $U(1)$ symmetry of the complex scalar field $L$.

The CTP formalism may be formulated in two ways, working either in the Heisenberg or interaction picture. In the former (see e.g.~\cite{Jordan:1986ug}), the density operator $\rho(\tilde{t}_i;\tilde{t}_i)$ does not evolve in time, remaining fixed at the boundary time $\tilde{t}_i$. As a result, the free propagators encode the initial conditions of the statistical ensemble, e.g.
\begin{equation}
[i\Delta^{N,\, 0}_{>}(x,y,0)]_\alpha^{\phantom \alpha \beta} \ =\ \braket{N_{\alpha}(x)N^{\beta}(y)}_{0}\ \equiv\ \frac{1}{Z} \, \mathrm{Tr}\,\rho(0) N_\alpha(x)N^\beta(y) \;.
\label{prop2}
\end{equation}
In this case, it is well-known that there does not exist a well-defined perturbative expansion. This may be understood by considering the Taylor expansion of the exponential decay to equilibrium: $
e^{-\Gamma t}\ = \ 1-\Gamma t+\frac{1}{2!}\big(\Gamma t\big)^2+\cdots$.
Any truncation of this expansion at a finite order in the decay rate $\Gamma$ leads to secular behaviour when $t>1/\Gamma$~\cite{Berges:2004yj}. This problem manifests in the Feynman-Dyson series of the Heisenberg interpretation of the CTP formalism as pinch singularities~\cite{Weldon:1991ek}, which result from ill-defined products of delta functions with identical arguments. As a consequence, it is necessary to work with dressed propagators, and therefore, ansaetze are required in order to extract particle number densities. The most common is the quasi-particle approximation known as the KB ansatz:
\begin{equation}
\label{eq:KBansatz}
[i\Delta_<^N(k,t)]_{\alpha}^{\phantom{\alpha}\beta} \ = \ 2\pi\delta(k^2-m_N^2)[n^N_{\mathrm{dress}}(\mathbf{k},t)]_{\alpha}^{\phantom{\alpha}\beta}\;,\qquad k^0>0\;,
\end{equation}
where $[n^N_{\mathrm{dress}}(\mathbf{k},t)]_{\alpha}^{\phantom{\alpha}\beta}$ approximates the matrix number density of spectrally-dressed particles.

On the other hand and in strong contrast to the earlier literature, it was shown recently~\cite{Millington:2012pf} that a perturbative framework of non-equilibrium thermal field theory is in fact viable, if we work instead in the interaction picture. Since the interaction-picture density operator $\rho(\tilde{t}_f;\tilde{t}_i)$ evolves in time, being evaluated at a macroscopic time $t=\tilde{t}_f-\tilde{t}_i$ after the specification of the initial conditions, the free positive-frequency Wightman propagator becomes
\begin{equation}
[i\Delta^{N, \, 0}_>(x,y,\tilde{t}_f;\tilde{t}_i)]_\alpha^{\phantom \alpha \beta} \ =\ \braket{N_{\alpha}(x;\tilde{t}_i)N^{\beta}(y;\tilde{t}_i)}_{t}\ \equiv\ \frac{1}{Z} \, \mathrm{Tr}\,\rho(\tilde{t}_f;\tilde{t}_i) N_\alpha(x;\tilde{t}_i)N^\beta(y;\tilde{t}_i)\;.
\label{prop1}
\end{equation}
In the heavy-neutrino mass eigenbasis and assuming spatial homogeneity, the free Wightman propagators then have the following explicit form:
\begin{align}
\label{Wfull}
&[i\widehat{\Delta}^{N,\,0}_{\gtrless}(k,k',\tilde{t}_f;\tilde{t}_i)]_{\alpha\beta}\ =\ 2\pi|2k_0|^{1/2}\delta(k^2-\widehat{m}_{N,\,\alpha}) \, 2\pi|2k_0'|^{1/2}\delta(k'^2-\widehat{m}_{N,\,\beta}) \, e^{i(k_0-k_0')\tilde{t}_f}\nonumber\\& ~
\times 
  \Big(\theta(\pm k_0)\theta(\pm k_0')\delta_{\alpha\beta} + [\theta(k_0)\theta(k_0')+\theta(-k_0)\theta(-k_0')][\widehat{n}^N(\mathbf{k},t)]_{\alpha\beta}\Big)(2\pi)^3\delta^{(3)}(\mathbf{k}-\mathbf{k}')\;,
\end{align}
depending on the time-dependent matrix number density $\bm{n}^N(\mathbf{k},t)$ of spectrally-free particles. These number densities appear as unknown functions in the Feynman-Dyson series, with their functional form being fixed only after the governing transport equations have been solved. Thus, the exponential decay to equilibrium is present implicitly in the free propagators of the theory, thereby avoiding the problem of secularity or pinch singularities, see~\cite{Millington:2012pf}.

Making a Markovian approximation and additionally setting $\widehat{m}_{N,\alpha}\simeq m_N=(\widehat{m}_{N,1}+\widehat{m}_{N,2})/2$, valid in the weakly-resonant regime, the free Wightman propagators in~\eqref{Wfull} reduce to
\begin{equation}
\label{homogenN}
[i \Delta^{N,\,0}_{\gtrless}(k,t)]_{\alpha}^{\phantom \alpha \beta}\ =\ 2\pi\delta(k^2-m_{N}^2)
  \Big(\theta(\pm k_0)\delta_{\alpha}^{\phantom{\alpha}\beta}\:+\:[n^N(\mathbf{k},t)]_{\alpha}^{\phantom \alpha \beta}\Big)\;,
\end{equation}
written here in a single-momentum representation and in a general flavour basis. The Markovian and homogeneous form of the \emph{free} propagator in~\eqref{homogenN} should be compared (for $k_0>0$) with the KB ansatz of the \emph{dressed} propagator in~\eqref{eq:KBansatz}, wherein we note that their spectral structure is identical in spite of the fact that the latter should be fully dressed spectrally.

In coordinate space, the KB equations take the following generic form (see e.g.~\cite{Prokopec:2003pj}):
\begin{align}
\label{KB1}
\Big(-\Box_x^2\:-\: |\bm{m}|^2\cdot\:+\: \bm{\Pi}_{\mathcal{P}}\ast\Big)\bm{\Delta}_{\gtrless} \ = \ -\:\frac{1}{2}\,\Big(\bm{\Pi}_>\ast\bm{\Delta}_<\:-\:\bm{\Pi}_<\ast\bm{\Delta}_>\:+\:2\,\bm{\Pi}_{\gtrless}\ast\bm{\Delta}_{\mathcal{P}}\Big) \;,
\end{align}
where $\Box_x^2\equiv\partial_{x^{\mu}}\partial_{x_{\mu}}$ is the d'Alembertian operator and $\cdot$ indicates matrix multiplication in flavour space. The $\ast$ denotes the convolution
\begin{equation}
\label{ast}
\bm{A}\ast\bm{B}\ \equiv\ \int_{z\,\in\,\Omega_t}\bm{A}(x,z,\tilde{t}_f;\tilde{t}_i)\cdot \bm{B}(z,y,\tilde{t}_f;\tilde{t}_i)\;,
\end{equation}
which is performed over the hypervolume $\Omega_t=[\tilde{t}_i,\tilde{t}_f]\times\mathbb{R}^3=[-\frac{t}{2},\frac{t}{2}]\times\mathbb{R}^3$, bounded temporally from below and above by the boundary and observation times, respectively~\cite{Millington:2012pf}.

The KB equation~\eqref{KB1} may be recast in a double momentum-space representation as follows:
\begin{align}
\label{KB2}
\Big(p^2\:-\: |\bm{m}|^2\cdot\:+\: \bm{\Pi}_{\mathcal{P}}\star\Big)\bm{\Delta}_{\gtrless} \ = \ -\:\frac{1}{2}\,\Big(\bm{\Pi}_>\star\bm{\Delta}_<\:-\:\bm{\Pi}_<\star\bm{\Delta}_>\:+\:2\,\bm{\Pi}_{\gtrless}\star\bm{\Delta}_{\mathcal{P}}\Big) \;.
\end{align}
Here, $\star$ denotes the weighted convolution integral
\begin{equation}
\label{star}
\bm{A}\star\bm{B}\ \equiv\ \int_{q,\,q'}\; (2\pi)^4\delta^{(4)}_t(q-q')\,\bm{A}(p,q,\tilde{t}_f;\tilde{t}_i)\cdot \bm{B}(q',p',\tilde{t}_f;\tilde{t}_i)\;,
\end{equation}
where
\begin{equation}
(2\pi)^4\delta^{(4)}_t(q-q')\ \equiv\ \int_{z\,\in\,\Omega_t}e^{-i(q-q')\cdot z}\ =\ (2\pi)^4\delta_t(q-q')\delta^{(3)}(\mathbf{q}-\mathbf{q}')
\end{equation}
and
\begin{equation}
\delta_t(q_0-q_0')\ \equiv\ \frac{1}{\pi}\frac{\sin[(q_0-q_0')t/2]}{q_0-q_0'}\;.
\end{equation}

By considering the Noether charge, see~\cite{Millington:2012pf}, the number density $\bm{n}(t,\mathbf{X})$ may be related unambiguously to the negative-frequency Wightman propagator via
\begin{equation}
\label{numdef}
\bm{n}(t,\mathbf{X})\ = \ \int^{(X)}_{p,\,p'}(p_0+p_0')\,i\bm{\Delta}_<(p,p',\tilde{t}_f;\tilde{t}_i)\;,\qquad \int^{(X)}_{p,\,p'}\ \equiv\ \int_{p,\,p'}e^{-i(p-p')\cdot X}\,\theta(p_0+p_0') \;,
\end{equation}
with $\int_p\equiv\int\!\frac{\D{4}{p}}{(2\pi)^4}$ and $\int_{p,p'}\equiv\int_p\int_{p'}$.
Following~\cite{Millington:2012pf}, we may then translate~\eqref{KB2} into the final rate equation for the number density
\begin{align}
\label{KB3}
&\frac{\D{}{\bm{n}(t,\mathbf{X})}}{\D{}{t}}\:-\:\int_{p,\,p'}^{(X)}(\mathbf{p}^2-\mathbf{p}'^2)\,\bm{\Delta}_{<}\:-\: \int_{p,\,p'}^{(X)}\Big([|\bm{m}|^2,\ \bm{\Delta}_{<}]\:-\:[\bm{\Pi}_{\mathcal{P}},\ \bm{\Delta}_{<}]_{\star}\Big) \nonumber\\&\qquad =\ -\:\frac{1}{2}\int_{p,\,p'}^{(X)}\Big(\{\bm{\Pi}_{>},\ \bm{\Delta}_{<}\}_{\star}\:-\:\{\bm{\Pi}_{<},\ \bm{\Delta}_{>}\}_{\star}\:+\:2\,[\bm{\Pi}_{<},\ \bm{\Delta}_{\mathcal{P}}]_{\star}\Big)\;,
\end{align}
where we use a compact notation
\begin{equation}
[\bm{A},\, \bm{B}]_{\star}\ \equiv\ \bm{A}\star\bm{B}\:-\:\bm{B}\star\bm{A}\;,\qquad
\{\bm{A},\, \bm{B}\}_{\star}\ \equiv\ \bm{A}\star\bm{B}\:+\:\bm{B}\star\bm{A}\;.
\end{equation}
The first two terms on the LHS of~\eqref{KB3} are the drift terms and the latter two account for mean-field effects, including oscillations; the terms on the RHS describe collisions. It should be stressed that~\eqref{KB3}, obtained without employing a gradient expansion or quasi-particle ansatz, is valid to any order in perturbation theory and accounts fully for spatial inhomogeneity, non-Markovian dynamics (memory effects) and flavour effects.

As identified in~\cite{Millington:2012pf}, the general rate equation in~\eqref{KB3} may be truncated in a perturbative loopwise sense in two ways: (i) spectrally: by truncating the external leg, we determine what is being counted, e.g.~inserting free propagators, we count spectrally-free particles; (ii) statistically: by truncating the self-energies, we determine the set of processes that drive the statistical evolution, e.g.~inserting one-loop self-energies, we include decay and inverse-decay processes.

\subsection{Heavy-neutrino rate equations}

Assuming spatial homogeneity and absorbing the principal part self-energy $\bm{\Pi}_{\mathcal{P}}^N$ into the thermal mass $\mat{M}_N^2 = |\mat{m}_N|^2 - \mat{\Pi}_\mathcal{P}^N$, we find the rate equation of the dressed heavy-neutrino number density
\begin{align}\label{eq:dressed_n}
&\frac{\D{}{}{\bm{n}}^{N}_{\mathrm{dress}}}{\D{}{t}} \ = \ \int^{(X)}_{k,\,k'}\: \bigg[ - \,i\, \big[\bm{M}^2_N,\ i \bm{\Delta}_<^N\big]\: - \:  \frac{1}{2}\Big( \big\{i \bm{\Pi}^N_<,\ i \bm{\Delta}^N_>\big\}_{\star}\:-\: \big\{i \bm{\Pi}^N_>,\ i \bm{\Delta}_<^N\big\}_{\star}\Big) \Bigg] \;.
\end{align}
Herein, we have also neglected the commutator involving $\mat{\Delta}^N_\mathcal{P}$ on the RHS of~\eqref{KB3}, since, in the weakly-resonant regime, it contains higher-order effects that are not relevant to this analysis.

Neglecting $\mathcal{O}(h^6)$ terms proportional to the lepton asymmetry, it is sufficient to approximate the charged-lepton and Higgs propagators, appearing in the heavy-neutrino self-energies, by their quasi-particle (narrow-width) equilibrium forms with vanishing chemical potential, i.e.
\begin{align}
\label{LPhiprops1}
i\Delta^{\Phi, \, \mathrm{eq}}_{\gtrless}(q)\ &=\ 2\pi\delta(q^2-M_{\Phi}^2) \, \big[\theta(\pm q_0)\:+\:n^{\Phi}_{\mathrm{eq}}(\mathbf{q}))\big]\;,\\
\label{LPhiprops2}
i\Delta^{L, \, \mathrm{eq}}_{\gtrless}(p)\ &=\ 2\pi\delta(p^2-M_L^2) \, \big[\theta(\pm p_0)\:+\:\theta(p_0)n^{L}_{\mathrm{eq}}(\mathbf{p})\:+\:\theta(-p_0)\overline{n}^{L}_{\mathrm{eq}}(\mathbf{p})\big]\;.
\end{align}
Here, $n^X_{\rm eq}(\ve p) = (e^{\sqrt{\mathbf{p}^2+M_X^2}/T} - 1)^{-1}$ is the Bose-Einstein distribution and $M_X$ is the thermal mass of species $X$. We are then left with the non-Markovian heavy-neutrino self-energies
\begin{align}
& [i\Pi_{\gtrless}^N(k,k'',\tilde{t}_f;\tilde{t}_i)]_{\alpha}^{\phantom{\alpha}\beta} \  =\ 2\,\widetilde{\Re}\,(h^\dag h)_{\alpha}^{\phantom \alpha \beta}\nonumber \\& \qquad \qquad \times \ \int_{p,\,q}(2\pi)^4\delta_t(k-p-q)\,(2\pi)^4\delta_t(k''-p-q)\,\Delta^{L,\mathrm{eq}}_\lessgtr(p) \, \Delta^{\Phi,\mathrm{eq}}_\lessgtr(q)\;.
\end{align}

We now perform a Wigner-Weisskopf approximation by replacing $\Omega_t$ by $\Omega_{\infty}$ in all spacetime integrals. In the double-momentum representation, this amounts to using the limit
\begin{equation}
\lim_{t\to\infty}\delta_t(k_0-p_0-q_0)\ =\ \delta(k_0-p_0-q_0)\;.
\label{appx1}
\end{equation}
At the same time, we replace $
e^{-i(k_0-k_0')\tilde{t}_f}\bm{\Delta}^N_{<}(k,k',\tilde{t}_f;\tilde{t}_i) \ \longrightarrow \ \bm{\Delta}^N_{<}(k,k',t)$, absorbing the free-phase evolution, which cancels that in the measure of~\eqref{numdef} in the energy-conserving limit. We then arrive at the Markovian rate equation for the dressed heavy-neutrino number density
\begin{align}\label{eq:KB_N_2}
\frac{\D{}{}[{n}^{N}_{\rm dress}]_{\alpha}^{\phantom{\alpha}\beta}}{\D{}{t}} \ &= \ \int_{k,\,k'} \theta(k_0+k_0')\: \bigg[ - \,i\, \big[M^2_N,  \, i \Delta_<^{N}(k,k',t)\big]_{\alpha}^{\phantom \alpha \beta} \notag\\ & \quad - \frac{1}{2} \, \Big(  [i \Pi^N_<(k)]_{\alpha}^{\phantom \alpha \gamma} \; [i \Delta^{N}_>(k,k',t)]_{\gamma}^{\phantom \gamma \beta} \; + \; [i \Delta^{N}_>(k,k',t)]_{\alpha}^{\phantom \alpha \gamma} \;  [i \Pi^N_<(k')]_{\gamma}^{\phantom \gamma \beta} \Big) \notag\\
& \quad + \frac{1}{2} \, \Big(  [i \Pi^N_>(k)]_{\alpha}^{\phantom \alpha \gamma} \; [i \Delta^{N}_<(k,k',t)]_{\gamma}^{\phantom \gamma \beta} \; + \; [i \Delta^{N}_<(k,k',t)]_{\alpha}^{\phantom \alpha \gamma} \;  [i \Pi^N_>(k')]_{\gamma}^{\phantom \gamma \beta} \Big) \bigg] \;,
\end{align}
where the Markovian self-energies
\begin{align}
i [\Pi^N_{\lessgtr}(k)]_{\alpha}^{\phantom \alpha \beta} \ &= \ 2\, \widetilde{\Re}(h^\dag h)_{\alpha}^{\phantom \alpha \beta} \, B_\lessgtr^{\mathrm{eq}}(k)
\end{align}
may be written in terms of the thermal loop functions
\begin{equation}
B^{\mathrm{eq}}_{\lessgtr}(k) \ \equiv \ \int_{p,\,q} \,(2 \pi)^4 \, \delta^{(4)}(p-k+q) \, \Delta_{\lessgtr}^{\Phi,\mathrm{eq}}(q) \, \Delta^{L,\mathrm{eq}}_{\lessgtr}(p) \;,\qquad B^{\mathrm{eq}}_{<}(-k) = B^{\mathrm{eq}}_{>}(k) \in \mathbb{R}\;.
\end{equation}
Subsequently, in the classical-statistical limit, these thermal loop functions can be written as
\begin{align}
B^{\mathrm{eq}}_{>}(k_0>0,\ve k) \ & = \ - \, \int \D{}{\Pi_\Phi} \int \D{}{\Pi_L} \, (2 \pi)^4 \, \delta^{(4)}(k-p_\Phi-p_L) \;, \label{eq:th_loop1}\\
B^{\mathrm{eq}}_{<}(k_0>0, \ve k) \ & = \ - \, \int \D{}{\Pi_\Phi} \int \D{}{\Pi_L} \, (2 \pi)^4 \, \delta^{(4)}(k-p_\Phi-p_L) \, n^\Phi_{\rm eq}(E_\Phi) \, n^L_{\rm eq}(E_L)\;. \label{eq:th_loop2}
\end{align}

Since we are interested in the asymmetry at $\mathcal{O}(h^4)$, we may truncate~\eqref{eq:KB_N_2} spectrally at zeroth loop order, replacing the external heavy-neutrino propagators by the free homogeneous propagator in~\eqref{homogenN}. The $k'$ integral in~\eqref{eq:KB_N_2} can then be performed trivially and we obtain the following rate equation for the spectrally-free number density $[n^N]_{\alpha}^{\phantom{\alpha}\beta}$:
\begin{align}\label{eq:KB_N_3}
&\frac{\D{}{}[{n}^{N}]_{\alpha}^{\phantom{\alpha}\beta}}{\D{}{t}} \ = \ \int_k \theta(k_0) \bigg\{ - \,i\, \big[M^2_N,  \, i \Delta_<^{N,\,0}(k,t)\big]_{\alpha}^{\phantom \alpha \beta} \nonumber\\&\qquad \qquad \qquad \qquad  - \;  \frac{1}{2} \, \Big( \big\{i \Pi^N_<(k),\, i \Delta^{N,\,0}_>(k,t)\big\}_{\alpha}^{\phantom \alpha \beta}\,-\, \big\{i \Pi^N_>(k),\, i \Delta_<^{N,\,0}(k,t)\big\}_{\alpha}^{\phantom \alpha \beta}\Big) \bigg\} \;.
\end{align}
After substituting for the explicit form of the free heavy-neutrino propagator given by~\eqref{homogenN}, we assume kinetic equilibrium, as described in~\cite{Dev:2014laa}, giving
\begin{equation}\label{eq:evol_N}
  \frac{\D{}{}[{n}^{N}]_{\alpha}^{\phantom{\alpha}\beta}}{\D{}{t}}
  \ = \ - \; i \,
  \Big[\mathcal{E}_N,\, n^{N}\Big]_\alpha^{\phantom \alpha \beta} \;
  + \; \Tdu{\big[\widetilde{\rm Re}
    (\gamma^{N,(0)}_{L \Phi})\big]}{}{}{\alpha}{\beta} \;
  - \; \frac{1}{2 \, n^N_{\rm eq}} \,
  \Big\{{n}^N, \, \widetilde{\rm Re}(\gamma^{N,(0)}_{L \Phi})
  \Big\}_{\alpha}^{\phantom{\alpha}\beta} \; .
\end{equation}
Here, we have defined the $\gCP$-``even'' rate 
$[\gamma^{N,(0)}_{L \Phi}]_{\alpha}^{\phantom{\alpha}\beta} \ \equiv \ \int_{N L \Phi} 2 \, h_\alpha h^\beta$,
where $h_{\alpha}$ are the tree-level Yukawa couplings. In addition, $\mat{\mathcal{E}}_{\!N}$ is the thermally-averaged effective energy matrix, given by~\eqref{thermen} evaluated with thermal masses and $g_{N}=1$.

Lastly, we separate the $\gCP$-``even'' and -``odd'' parts of~\eqref{eq:evol_N}, giving
\begin{subequations}
  \label{nfin}
\begin{align}
  \frac{\D{}{}[\underline{n}^{N}]_{\alpha}^{\phantom{\alpha}\beta}}{\D{}{t}}
  \ & = \ - \; \frac{i}{2} \,
  \Big[\mathcal{E}_N,\, \delta n^{N}\Big]_\alpha^{\phantom \alpha \beta} \;
  + \; \Tdu{\big[\widetilde{\rm Re}
    (\gamma^{N,(0)}_{L \Phi})\big]}{}{}{\alpha}{\beta} \;
  - \; \frac{1}{2 \, n^N_{\rm eq}} \,
  \Big\{\underline{n}^N, \, \widetilde{\rm Re}(\gamma^{N,(0)}_{L \Phi})
  \Big\}_{\alpha}^{\phantom{\alpha}\beta}
   \;, \label{eq:evol_n}\\[3pt]
  \frac{\D{}{[\delta n^N]_\alpha^{\phantom \alpha \beta}}}{\D{}{t}} \ &
  = \ - \; 2 \, i \, 
  \Big[\mathcal{E}_N,\, \underline{n}^{N}\Big]
  _\alpha^{\phantom \alpha \beta} \; - \; \frac{1}{2 \, n^N_{\rm eq}}  \,
  \Big\{\delta n^N, \, \widetilde{\rm Re}(\gamma^{N,(0)}_{L \Phi})
  \Big\}_{\alpha}^{\phantom{\alpha}\beta}\;. \label{eq:evol_dn} 
\end{align}
\end{subequations}
These final heavy-neutrino rate equations agree, up to $\mathcal{O}(h^4)$ as considered here, with those obtained by the semi-classical approach [cf.~\eqref{eq:evofinal2}--\eqref{eq:evofinal3}].

\subsection{Source term for the asymmetry}

In this section, we describe the explicit form of the dressed negative-frequency heavy-neutrino Wightman propagator, as derived in~\cite{Dev:2014wsa} using the perturbative approach of~\cite{Millington:2012pf}, making comparison with the standard quasi-particle or KB ansatz. In so doing, we will illustrate that both heavy-neutrino mixing and oscillations provide distinct contributions to the $\mathcal{O}(h^4)$ lepton asymmetry in the weakly-resonant regime and that the contribution of flavour mixing is discarded when the standard quasi-particle approximations are employed.

\begin{figure}
\centering
\vspace{-3em}
\begin{align*}
\raisebox{-1.15em}{\includegraphics[scale=0.7]{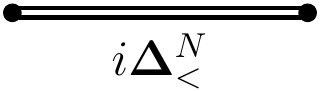}}\ = \ \raisebox{-1.15em}{\includegraphics[scale=0.7]{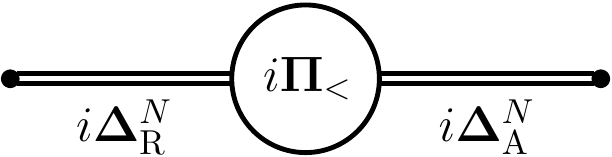}}\ +\ \raisebox{-1.2em}{\includegraphics[scale=0.7]{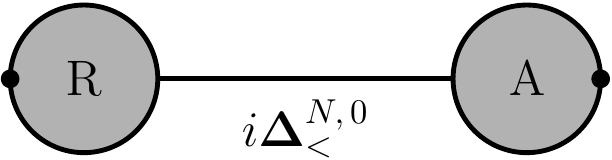}}
\end{align*}
\caption{Diagrammatic representation of the dressed negative-frequency heavy-neutrino matrix Wightman propagator $i\bm{\Delta}_<^{N}$. Double lines indicate fully-dressed propagators and single lines, propagators dressed with only dispersive corrections. Unshaded blobs are the relevant self-energies and shaded blobs are the amputated self-energy corrections to the vertices.\label{fig:prop}}
\end{figure}

Working again in the Markovian regime and assuming that the charged-lepton and Higgs propagators have the equilibrium forms in~\eqref{LPhiprops1} and~\eqref{LPhiprops2}, the dressed negative-frequency heavy-neutrino Wightman propagator is determined by the following Schwinger-Dyson equation~\cite{Millington:2012pf}:
\begin{align}\label{eq:SD}
&i \mat{\Delta}^{N}_{<}(k,k',t) \ = \ i \mat{\Delta}^{N,\, 0}_{<}(k,k',t)\:+\: i \mat{\Delta}^{N,\, 0}_{\mathrm{R}}(k) \cdot i \mat{\Pi}_<(k)(2\pi)^4\delta^{(4)}(k-k') \cdot  i \mat{\Delta}^{N}_{\mathrm{A}}(k')\nonumber\\&\qquad+\: i \mat{\Delta}^{N,\, 0}_{\mathrm{R}}(k) \cdot i \mat{\Pi}_{\mathrm{R}}(k) \cdot  i \mat{\Delta}^{N}_{<}(k,k',t) \:+\: i \mat{\Delta}^{N,\, 0}_{<}(k,k',t) \cdot i \mat{\Pi}_{\mathrm{A}}(k') \cdot  i \mat{\Delta}^{N}_{\mathrm{A}}(k') \;.
\end{align}
In addition, the Schwinger-Dyson equation for the advanced propagator has the closed form
\begin{equation}
\label{eq:SDA}
i \mat{\Delta}^{N}_{\mathrm{A}}(k) \ = \ i \mat{\Delta}^{N,\, 0}_{\mathrm{A}}(k) \:+\: i \mat{\Delta}^{N,\, 0}_{\mathrm{A}}(k) \cdot i \mat{\Pi}_{\mathrm{A}}(k) \cdot  i \mat{\Delta}^{N}_{\mathrm{A}}(k) \;.
\end{equation}
As shown in~\cite{Dev:2014wsa},~\eqref{eq:SD} and~\eqref{eq:SDA} can be solved iteratively to give an explicit form for the dressed negative-frequency Wightman propagator, shown diagrammatically in Figure~\ref{fig:prop}:
\begin{align}
\label{fullWight}
&[i\Delta^{N}_{<}(k,k',t)]_{\alpha}^{\phantom{\alpha}\beta}\ =\  [i\Delta^N_{\mathrm{R}}(k)]_{\alpha}^{\phantom{\alpha}\gamma}[i\Pi^N_<(k)]_{\gamma}^{\phantom{\gamma}\delta}(2\pi)^4\delta^{(4)}(k-k')[i\Delta^N_{\mathrm{A}}(k')]_{\delta}^{\phantom{\delta}\beta}\nonumber\\&\quad +\:\sum_{m\:=\:0}^{\infty}\Big[\big(i\Delta^0_{\mathrm{R}}(k)\cdot i\Pi^N_{\mathrm{R}}(k)\big)^{m}\Big]_{\alpha}^{\phantom{\alpha}\gamma} \, [i\Delta_<^{N,\,0}(k,k',t)]_{\gamma}^{\phantom{\gamma}\delta}\sum_{n\:=\:0}^{\infty}\Big[\big(i\Pi^N_{\mathrm{A}}(k')\cdot i\Delta^{N,\,0}_{\mathrm{A}}(k') \big)^n\Big]_{\delta}^{\phantom{\delta}\beta}\;.
\end{align}

The first term on the RHS of~\eqref{fullWight} contributes to the washout from $\Delta L = 0$ and $\Delta L = 2$ scatterings. The second term instead contributes to the source term for the asymmetry. Rotating first to the mass eigenbasis, it was shown in~\cite{Dev:2014wsa} that this contribution may be written in terms of the resummed Yukawa couplings $\mathbf{h}^{\alpha}$~\cite{Pilaftsis:2003gt, Dev:2014laa} by virtue of the equivalence
\begin{align}
&\widehat{h}^\alpha \bigg[ \sum_{n\:=\:0}^\infty \Big ( i \widehat{\Delta}_{\mathrm{R}}^0(k) \cdot i \widehat{\Pi}_{\mathrm{R}}(k) \Big)^n \bigg]_{\alpha}^{\phantom \alpha \beta} \ \sim \ \widehat{\mathbf{h}}^\beta \;.
\label{eq:diag}
\end{align}
In addition, the same contribution can be recast as
\begin{align}
\label{fullWight2}
&[i\widehat{\Delta}^{N}_{<}(k,k',t)]_{\alpha \beta}\ \supset  \ [\widehat{\Delta}_{\mathrm{R}}^N(k)]_{\alpha \gamma} \, \Big( [\widehat{\Delta}_{\mathrm{R}}^{N,\,0}(k)]^{-1}_{\gamma \gamma}  \, [i \widehat{\Delta}_<^{N,\,0}(k,k',t)]_{\gamma \delta} \, [\widehat{\Delta}_{\mathrm{A}}^{N,\,0}(k')]^{-1}_{\delta \delta}\Big) \, [\widehat{\Delta}_{\mathrm{A}}^N(k')]_{\delta \beta}\;.
\end{align}
On the other hand, the flavour-covariant KB ansatz for the heavy-neutrino propagator is
\begin{equation}
[i\widehat{\Delta}^{N}_{\mathrm{KB},\,<}(k,k',t)]_{\alpha\beta}\ =\ 2\pi\delta(p^2-\widehat{M}^2_{N,\,\alpha}) \, 2\pi\delta(k'^2-\widehat{M}^2_{N,\,\beta}) \, [n^N_{\mathrm{KB}}(\mathbf{k},t)]_{\alpha\beta} \, (2\pi)^3\delta^{(3)}(\mathbf{k}-\mathbf{k}')\;,
\end{equation}
where we have restricted $k_0>0$. It is then clear that the KB ansatz satisfies
\begin{align}
\label{KBzero}
\big(k^2-\widehat{M}^2_{N,\,\alpha}\big)\, [i\widehat{\Delta}^{N}_{\mathrm{KB},\,<}(k,k',t)]_{\alpha\beta}\ =\ 0\;,\qquad [i\widehat{\Delta}^{N}_{\mathrm{KB},\,<}(k,k',t)]_{\alpha\beta}\,\big(k'{}^2-\widehat{M}^2_{N,\,\beta}\big)\ =\ 0\;,
\end{align}
whereas the dressed heavy-neutrino Wightman propagator in~\eqref{fullWight} does not, due to the mixing that gives rise to the resummed Yukawas. We therefore conclude that the KB ansatz discards the phenomenon of flavour mixing, accounting only for the separate phenomenon of oscillations.

\begin{figure}
\begin{equation*}
\parbox{11em}{\includegraphics[width = 11em]{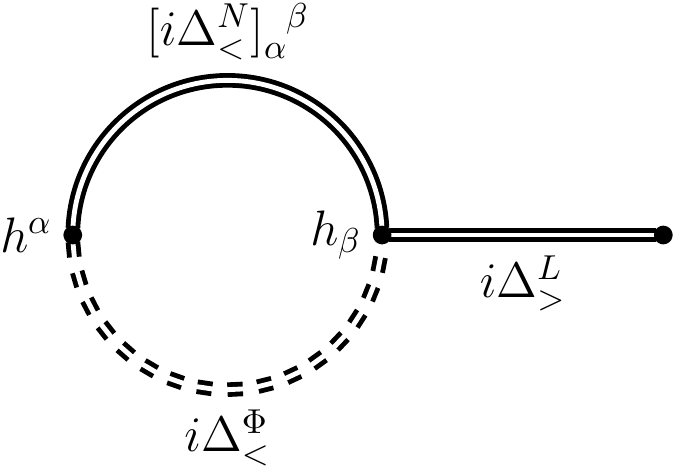}} \ \supset \ \parbox{11em}{\includegraphics[width = 11em]{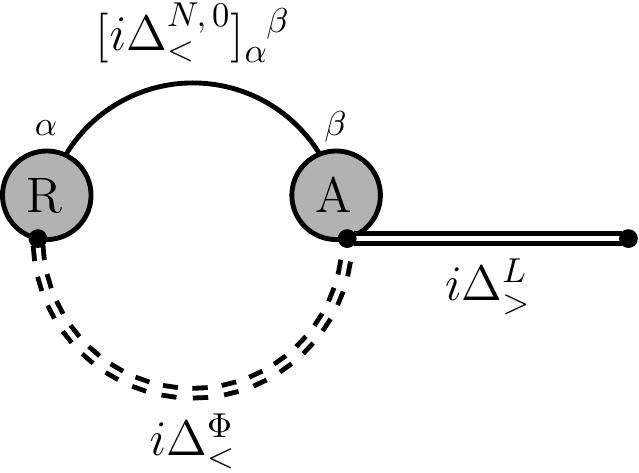}} \ \simeq \ \parbox{11em}{\includegraphics[width = 11em]{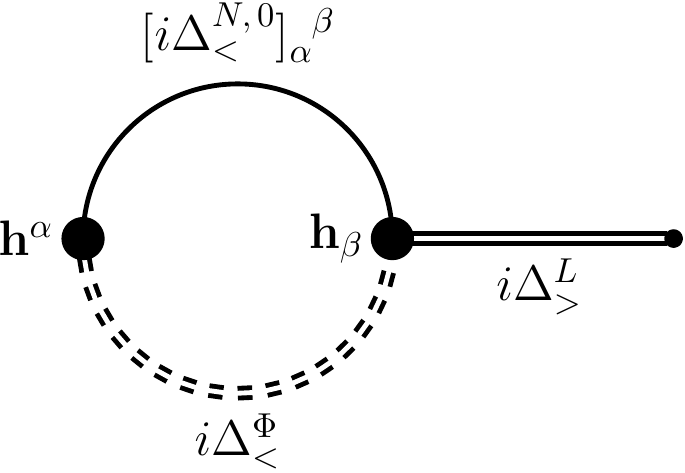}} 
\end{equation*}
\caption{Diagrammatic representation of the factorization of absorptive transitions into the resummed Yukawa couplings in the source term for the charged-lepton asymmetry. \label{fig:self}}
\end{figure}

Using the equivalence in~\eqref{eq:diag}, the effect of mixing, arising from absorptive transitions in the dressed propagator~\eqref{fullWight}, can be absorbed into the resummed Yukawa couplings. Once this is done, it is then appropriate to replace the non-homogeneous free heavy-neutrino propagator $\mat{\Delta}_<^{N,0}(k,k',t)$, appearing on the RHS of~\eqref{fullWight}, by its homogeneous counterpart, as given by~\eqref{homogenN}. This procedure is illustrated diagrammatically in Figure~\ref{fig:self}. We then find that the contribution from the charged-lepton self-energy to the source term for the asymmetry is given by 
\begin{align}
\label{eq:source_res_Yuk}
\frac{\D{}{\delta n^L} }{\D{}{t}} \  &\supset \ -\int_k \theta(k_0) \Big[ \hrs{}{\beta} \hr{}{\alpha} \Big([i\Delta^{N,\,0}_<(k,t)]_{\alpha}^{\phantom \alpha \beta} \, B_{>}^{\mathrm{eq}}(k)- [i\Delta^{N,\,0}_>(k,t)]_{\alpha}^{\phantom \alpha \beta} \, B_{<}^{\mathrm{eq}}(k) \Big)-\gC\!. c. \Big] \;.
\end{align}
We again assume kinetic equilibrium and separate out the $\gCP$-``even'' and ``-odd'' parts of the heavy-neutrino number density, i.e.~$\mat{\underline{n}^{N}}$ and $\mat{\delta n^N}$, giving the final form
\begin{equation}
\label{eq:evol_dL}
\frac{\D{}{\delta n^L}}{\D{}{t}} \ 
  = \ 
  \bigg(\frac{[\underline{n}^{N}]_{\alpha}^{\phantom{\alpha}\beta}}
  {n^N_{\rm eq}} \, - \, \delta_{\alpha}^{\phantom{\alpha}\beta}\bigg) \,
  \Tdu{[\delta \gamma^{N}_{L \Phi}]}{}{}{\beta}{\alpha} \;
  + \; \frac{[\delta n^N]_{\alpha}^{\phantom{\alpha}\beta}}{2\,n^N_{\rm eq}} \,
  \Tdu{[\gamma^{N}_{L \Phi}]}{}{}{\beta}{\alpha} \; + \; W[\delta n^L]\;,
\end{equation}
where $W[\delta n^L]$ denotes washout terms not studied explicitly here. The source term in~\eqref{eq:evol_dL} agrees with~\eqref{eq:evofinal1}, as derived in the semi-classical approach in~\cite{Dev:2014laa}, up to ${\cal O}(h^4)$. Evidently, both flavour mixing and oscillations can be identified in~\eqref{eq:evol_dL} as providing distinct contributions to the final asymmetry.

\begin{figure}
\begin{align*}
\Delta^{0,-1}_N \; \star \; \raisebox{0.3em}{\parbox{5em}{\includegraphics[width=5em]{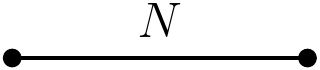}}} \quad &\sim \quad \parbox{9.5em}{\includegraphics[width=9.5em]{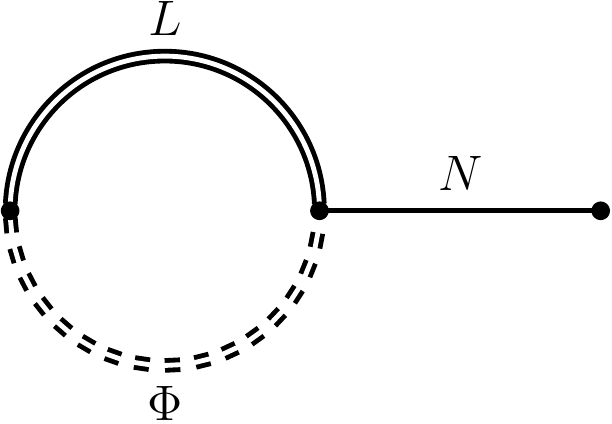}}\\[6pt]
\Delta^{0,-1}_L \; \star \; \raisebox{0.3em}{\parbox{5em}{\includegraphics[width=5em]{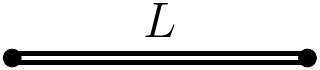}}} \quad &\sim \quad \parbox{9.5em}{\includegraphics[width=9.5em]{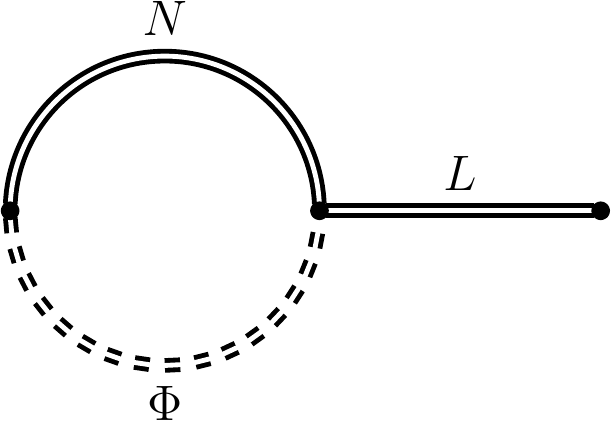}} \quad \supset \quad \parbox{9.5em}{\includegraphics[width=9.5em]{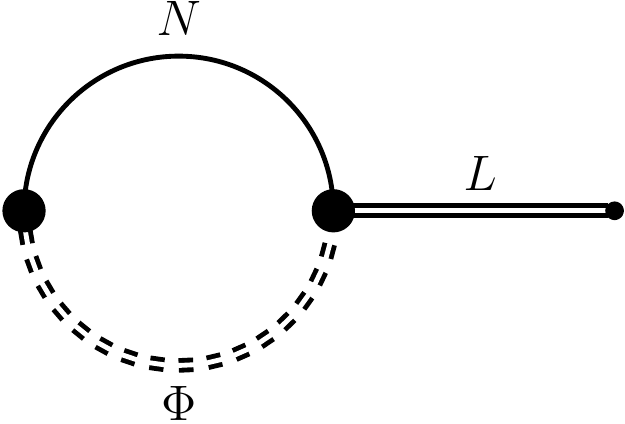}}
\end{align*}
\caption{Diagrammatic representation of the truncation procedure, proceeding \emph{spectrally} for the heavy-neutrino rate equation (top) and \emph{statistically} for the charged-lepton asymmetry (bottom). \label{fig:trunc}}
\end{figure}

The self-consistent loopwise perturbative truncation scheme described in this section is summarized diagrammatically in Figure~\ref{fig:trunc}. Before concluding, however, it is important to stress the following. In the weakly-resonant regime, while it is appropriate to truncate the heavy-neutrino rate equations \emph{spectrally} at zeroth order, inserting free heavy-neutrino propagators in the external legs, it is \emph{not} appropriate to insert the same free heavy-neutrino propagators in the charged-lepton self-energies of the source term for the asymmetry. The latter would instead correspond to a zeroth-order \emph{statistical} truncation, which, were it performed, would have the same impact as making a KB ansatz, discarding the physical phenomenon of flavour mixing.

\section{Conclusions}\label{sec:conclusion}

We have presented a novel approach to the study of flavour effects in RL by means of a {\it fully} flavour-covariant formalism for transport phenomena~\cite{Dev:2014laa}. Our {\em manifestly} flavour-covariant rate equations for heavy-neutrino  and
 lepton number densities provide a complete and unified
description of  RL, capturing  three distinct physical  effects: (i)
resonant  mixing between the heavy-neutrino  flavours, (ii) coherent
oscillations  between  different heavy-neutrino  flavours and  (iii)
quantum  decoherence   effects  in  the   charged-lepton  sector. The full impact of these flavour off-diagonal effects has been illustrated in an RL$_\tau$ model, where the total lepton asymmetry varies as much as an order of magnitude between the partially flavour off-diagonal treatments.

We have also presented an embedding of our flavour-covariant formalism~\cite{Dev:2014laa} within a perturbative formulation of non-equilibrium thermal field theory~\cite{Millington:2012pf}, enabling us to extract physically-meaningful particle number densities at any order in perturbation theory, without the need for quasi-particle ansaetze. In this novel quantum field-theoretic approach to the KB formalism~\cite{Dev:2014wsa}, we have justified, at leading order and in the weakly-resonant regime, the semi-classical Boltzmann approach adopted in~\cite{Dev:2014laa}, capturing all flavour effects pertinent to RL. In particular, we have confirmed that the mixing and oscillations between different heavy-neutrino flavours are two {\em physically-distinct} phenomena. We emphasise that the former effect has been implicitly disregarded in previous KB studies that rely on particular quasi-particle ansaetze.  

\ack

The work  of P.S.B.D. and  A.P.  is supported  by the
Lancaster-Manchester-Sheffield  Consortium   for  Fundamental  Physics
under  STFC   grant  ST/L000520/1. The work of P.M. is supported by a University Foundation Fellowship (TUFF) from the Technische Universit\"{a}t M\"{u}nchen and the Deutsche Forschungsgemeinschaft (DFG) cluster of
excellence Origin and Structure of the Universe. The work of D.T. has been supported by a fellowship of the EPS  Faculty of  the  University  of Manchester.

\end{document}